# A liquid Xenon Positron Emission Tomograph for small animal imaging : first experimental results of a prototype cell.

M.-L. Gallin-Martel[a*1], L. Gallin-Martel[a], Y. Grondin[b], O. Rossetto[a], J. Collot[a], D. Grondin[a], S. Jan[a2], Ph. Martin[a], F. Mayet[a], P. Petit[a], F. Vezzu[a]

[a] *Laboratoire de Physique Subatomique et de Cosmologie, Université Joseph Fourier Grenoble 1, CNRS/IN2P3, Institut National Polytechnique de Grenoble, 53 avenue des Martyrs, 38026 Grenoble cedex, France*

[b] *Laboratoire TIMC/IMAG, CNRS et Université Joseph Fourier, Pavillon Taillefer, 38706 La Tronche cedex, France*

**Abstract**

A detector using liquid Xenon (LXe) in the scintillation mode is studied for Positron Emission Tomography (PET) of small animals. Its specific design aims at taking full advantage of the Liquid Xenon scintillation properties. This paper reports on energy, time and spatial resolution capabilities of the first LXe prototype module equipped with a Position Sensitive Photo-Multiplier tube (PSPMT) operating in the VUV range (178 nm) and at 165 K. The experimental results show that such a LXe PET configuration might be a promising solution insensitive to any parallax effect.

---

[1] Corresponding author Tel +33 4 76 28 41 28; Fax : +33 4 76 28 40 04
E-mail adress : mlgallin@lpsc.in2p3.fr[1]

[2] Present address : Service Hospitalier Frédéric Jolliot (SHFJ), CEA, F91401 Orsay, France

## 1 Introduction.

Positron Emission Tomography (PET) is a non-invasive clinical and research imaging technique in nuclear medicine using radio-labeled molecules to image molecular biological processes in vivo. It is also a powerful tool in modern clinical applications for cancer diagnosis. Tremendous experimental efforts on a host of techniques have been made in the field of PET imaging, in particular towards the development of new generation high resolution PET cameras dedicated to small animal imaging [1,4]. However, improvements are still needed with respect to the spatial resolution and sensitivity of the technique for its application to specific human organs and in particular to small animals. Recent technological advances have led to the development of dedicated small animal scanners which can be a tool to measure the kinetics of labeled drugs. The aim is to study interactions between novel drugs and their biological target, metabolism and morphological changes in animal models of disease. High time and energy resolution are required as well as DOI measurement for new generation PET prototypes.

Standard PETs are made of several longitudinal slices of radially-oriented crystals leading to the main drawback of such a geometry : when measuring the spatial coordinates of each interacting gamma, while two coordinates are well measured owing to the transverse granularity of the device, the Depth Of Interaction (DOI) cannot be determined with an equivalent accuracy. The hit is then reconstructed at the foot of the crystal (see Fig. 1). The consequence of this parallax effect is therefore a degradation of the resolution in the image reconstruction for positron annihilation off the PET longitudinal axis. Typically, for a microPET FOCUS 220, with point like $^{18}F$ sources, the radial resolution increases from ~1.3 mm to ~3.5 mm when the source is shifted from 0 to 8 cm off the axis [5].

We describe here a novel design of PET scanner for small animals using the scintillation properties of the Liquid Xenon in an axial geometry to provide a three dimensional gamma reconstruction free of parallax error. Liquid Xenon is well known to be an excellent detecting medium for γ-rays due to its high atomic number and density. When compared to commonly used crystals (see Tab. 1), it appears that its decay time is short : 3 to 30 ns [6,7], the spread being due to the various scintillation modes of the Xe excimers. Its fast scintillation light yield is comparable to that of NaI. Several groups have been studying the possibility to use Liquid Xenon (LXe) as a sensitive medium in PET imaging field, either in a charge collecting mode [11] or in scintillation mode [12-15], or for other applications (e.g Dark Matter search) [16-18].

In the proposed PET design, the Xenon is contained in aluminum light guides read at each end with Position Sensitive PhotoMultiplier Tubes (PSPMT). When designing a LXe-based PET camera, there are at least three experimental difficulties to overcome:

- the purity of Xenon, where contamination should not exceed a few ppm in scintillation mode,
- the operating conditions at cryogenic temperature (typically 165 K),
- the detection of UV photons around 178 nm [19].

A LXe PET prototype module has been built and simulated using the GEANT4 [20] Monte Carlo code linked to the GATE [21] package adapted to this new scanner geometrical configuration. We first report in this paper on the z (axial) coordinate analytical calculation. Various models used by other collaborations that also attempt to solve the DOI problem with a 3D axial geometry detector are discussed. Then a comparison is done between simulation and experimental data that correspond to our first characterization measurements of the prototype module in terms of energy, time and spatial resolution capabilities which are three relevant parameters for the optimization of the proposed LXe μPET prototype.

## 2 Experimental set-up.

The active volume of this prototype LXe μPET camera is a ring featuring an internal diameter of 8 cm and a radial extension of 25 mm (see Fig. 2). This module is placed in a cryostat composed of aluminum walls and filled with Liquid Xenon. Sixteen identical modules of the type shown on Fig. 3 are immersed in this ring. Each module presents a 2 x 2 $cm^2$ cross-section in the transaxial plane of the camera. The axial field of view is 5 cm. A module is optically subdivided into 40 $MgF_2$-coated aluminum UV light guides, each featuring a 5 x 2 $mm^2$ cross-section. The scintillation photons propagate in the guide by specular reflection. They are collected on both ends of each module by two PSPMT (see Fig. 4). The (x,y) positions measured by the PSPMT determine which light guides have been fired. The light attenuation which depends on the guide reflectivity allows the axial coordinate (z) to be deduced from dynode signals asymmetry.

This experimental set-up used for the test of the first prototype module is described on Fig. 5. The Xenon is grade 4.8 (i.e. a purity better than 20 ppm), which is pure enough for the detection of VUV photons after a few centimetre path length, whereas a maximum contamination of a few ppb is required if the aim is to drift ionization charges in an electric field [22]. Following the layout displayed on Fig. 5, Xenon is liquefied in the compressor (the liquid Xe compressor was purchased from Air Liquide DTA), then transferred to a container inside the cryostat. The temperature inside the cryostat is kept around 165 K via a liquid nitrogen heat exchanger. The cryostat is flushed with argon gas.The temperature fluctuation is less than a few tenths of a degree. The Xenon container is a stainless steel cylinder 50 mm long and 40 mm in diameter, closed at each end with a 3 mm thick suprasil window. This container can house various types of cells (see Fig. 4). An experimental test bench has been built to allow the x and y measurement and furthermore to evaluate the z-axis localization, the energy and time resolutions. A $^{22}$Na source is mounted on a small carriage moving along the z-axis, underneath the cryostat. The collimation is done by two pairs of lead collimators placed respectively at the top and bottom of the $^{22}$Na source inducing a 2 mm gap (see Fig. 5). A LYSO crystal coupled to a photomultiplier tube completes the experimental set-up providing the coincidence signal. The VUV photons are collected with a PSPMT at each end. In the present set-up the PSPMTs are not immersed in the LXe and a thin argon gap (0.3 mm) is kept between the cell quartz window and the PMT entrance window. Photons then encounter refraction indices of 1.69 (Liquid Xenon [23]), 1.58 (cell quartz window), 1.0 (gas), and finally 1.58 (PSPMT entrance quartz window) leading to a 36 degree limit angle (LA). PSPMTs with the required specifications, i.e high Quantum Efficiency (QE) at 178 nm (QE = 20 %) and still working at low temperatures (165 K), are not commercially available yet. Hamamatsu provided us with two prototype tubes, belonging to the R8520-06-C12 series [24], having five aluminum strips deposited on their window to reduce the resistivity of the photocathode at 165 K. Their QE was around 20 %. On the whole however, these tubes were very satisfactory for exploring the feasibility of a PET based on the Liquid Xenon scintillation.

The read-out electronics (see Fig. 6) operates at room temperature and is composed of standard NIM and CAMAC modules. The charge measurement of PSPMT anode signals (6x and 6y) is performed using gated integrators and 12-bit Analogue to Digital Converters (QDC). The last dynode signal of each PSPMT is splitted in order to be treated like anode signals and also to be used for time tagging and experimental set-up triggering. The signals provided by the PMT coupled to the LYSO crystal are processed in the same way. The time tagging and triggering electronics comprises: two Constant Fraction Discriminators (CFD), a coincidence module and a Time to Digital Converter (TDC). One CFD processes the LYSO PMT dynode signal, the second one is fed with the analog sum of the two PSPMT dynode signals. The coincidence module uses CFD output signal to trigger the TDC, the integrators and the VME-based data acquisition.

## 3 The Monte Carlo LXe μPET camera module simulation.

In order to simulate the LXe μPET camera module, the GATE toolkit has been used [21,25]. Each module was created as a square section box divided into 40 smaller boxes of rectangular section representing the light guides. The LXe material was defined in the material database file of GATE and assigned to the module. The aluminum coats covering the light guides have not been simulated in GATE, their influence on the detection of the 511 keV gamma photons being considered negligible [14,26]. However, UV photon transport in the aluminum guides and detection resolution were simulated with a custom MC code. The code is able to track individual scintillation photons from their emission point along the aluminum guide to the photodetectors, taking into account reflection as well as absorption. However it does not include the photon detection process operated by the PMTs. The source was a hypothetical 511 keV gamma photon emitter placed at the real distance below the detector. It was a 1 million Becquerel sphere of 1 mm radius. The lead shields of the experiment were also simulated. The presence of the LYSO acting as a trigger or a "collimator" was taken into account by simulating additional lead shields such that, combined to the other shields, the gamma photons are collimated through a square hole of the same area as the LYSO's sensitive surface. The output data of the GATE simulations consisted in the collection of all interactions occurring in the detector and in the other simulated components of the set-up. The most important features that are used for simulating the generation of the UV photons and therefore the detection in our Monte Carlo code are the amount of deposited energy and the localization of the hits in the detector module.

The values of the simulation parameters are:

- light guide of 50 mm long and 5 x 2 $mm^2$ in cross-section
- 4 guides in x and 10 in y
- light guide reflectivity CR=0.8
- limit angle LA=36°
- PSPMT quantum efficiency QE=20 %
- LXe refractive index n=1.69 [23]
- scintillation yield in LXe : 46 $10^3$ UV/MeV [8]

## 4 Light guide modelling and simulation result.

### *4.1 Light guide modeling.*

Experimental studies on DOI measurement have been already carried out using axial scintillator arrangement [27-30]. The DOI quantification described in [27,28] is based on a linear approach of the light attenuation in the medium. The simulation shows that this model can only be applied with short crystals (2-3 cm). Large discrepancies are induced if longer scintillators are concerned. A second approach described in [29] introduces an angle θ defined as arctan(T/B), where T and B are the signals collected on both scintillator ends. The DOI calibration of the crystal is then derived from a polynomial fit of the function DOI=f(θ). In this example no light propagation model is introduced but the polynomial fit insures a reliable DOI calibration even for long crystals. Another approach [30] (CIMA collaboration) assumes that the amount of light collected at a crystal end decays exponentially with the axial z coordinate of the $\gamma$ photon interaction point. In the present device, the attenuation length $\lambda_{ref}$ of the light guide can be introduced. If we assume that the intrinsic attenuation length $\lambda_{int}$ of the LXe is much larger than $\lambda_{ref}$, the light attenuation in the detector can be described by:

$$I(z) = I_0 e^{\frac{-z}{\lambda_{ref}}} \qquad (1)$$

The attenuation length $\lambda_{ref}$ depends on the light guide geometry and its reflectivity. It also depends on the limit angle LA : small values of LA favours photons which undergo small numbers of reflections consequently increasing the $\lambda_{ref}$ value. The number of photoelectrons collected at each module end is given by:

$$N_l(z) = A \cdot \frac{N_0}{2} \cdot \exp\left(\frac{-l}{\lambda_{ref}}\right) \cdot \exp\left(\frac{-z}{\lambda_{ref}}\right) \qquad (2)$$

$$N_r(z) = A.\frac{N_0}{2} \cdot \exp\left(\frac{-l}{\lambda_{ref}}\right) \cdot \exp\left(\frac{z}{\lambda_{ref}}\right)$$

Where :

- the origin of the z axial coordinates is taken at the centre of the module,
- A is a constant depending on LA and on the PMT quantum efficiency,
- $N_0$ is the amount of scintillation photons,
- l is the module half length.

In Fig 7 the total amount of photoelectrons collected at each module end is plotted, scanning the total length of the simulated LXe module in steps of 5 mm. The simulated data are well fitted with an exponential curve.

*4.2 The LXe pollution modeling in the simulation.*

The exponential approach is very attractive because light propagation and collection can be described with simple analytic expressions, and moreover the propagation medium (light guide) is fully characterized by a single parameter $\lambda_{ref}$. However, the attenuation length of LXe strongly depends on its purity. In the past, various experiments were dedicated to the measurement of this parameter [23, 31-33]. It was shown that as long as pure Xenon is concerned its intrinsic attenuation length can be neglected when compared to the $\lambda_{ref}$ obtained with the currently studied LXe μPET module. The absorption of Xenon scintillation light by impurities (mainly $H_2O$ and $O_2$) is described by the relation:

$$I(z) = I_0 \cdot \exp\left(\frac{-z}{\lambda_{abs}}\right) = I_0 \cdot \exp(-\sigma\rho N z) \quad (3)$$

where $\sigma$ is the cross section of the interaction, $\rho$ is the impurities concentration and N the LXe density. The equivalent attenuation length is given by:

$$\lambda_{abs} = \frac{1}{\sigma\rho N} \quad (4)$$

This attenuation length corresponds to the $\lambda_{bulk}$ parameter mentioned in [34]. The effective attenuation length (due to impurities) $\lambda_{eff}$ can be obtained in the same way and finally the overall $\lambda$ is given by:

$$\frac{1}{\lambda} = \frac{1}{\lambda_{ref}} + \frac{1}{\lambda_{eff}} \quad (5)$$

This relation shows that exponential model allows the LXe pollution to be easily introduced in the analytic expressions that describe light propagation.

### *4.3 Determination of the attenuation length $\lambda ref$.*

The first step to perform DOI measurement is to qualify the light propagation medium. The light guide attenuation length $\lambda_{ref}$ can be derived from Eq. (2):

$$\ln\left(\frac{N_r}{N_l}\right) = \frac{2z}{\lambda_{ref}} \qquad (6)$$

In Fig. 8 the expression $\ln(N_r/N_l)$ is plotted for a 5 mm step module scanning. It can be seen that the exponential model is very satisfactory at this stage. The $\lambda_{ref}$ value is found to be equal to 26.5 ± 0.1 mm. In this simulation the $^{22}Na$ source is perfectly collimated and no $\Delta z$ spread is introduced in the z coordinate of interaction for a given source position. This leads to a linear trend even for $z_{source}$=+/-25 mm which corresponds to the module edges. A small deficit of UV photons is observed on the module edges when the simulation takes into account a "beam width".

### *4.4 Determination of the interaction point.*

The reconstructed z coordinate is derived using Eq.(7).

$$z_{rec} = \frac{1}{2}\lambda_{\mathrm{ref}} \ln\left(\frac{N_r}{N_l}\right) \qquad (7)$$

The distribution of the reconstructed $z_{rec}$ position for a source located at $z_{source}= 0$ is shown on Fig. 9. A Gaussian fit of this distribution for different values of $z_{source}$ provides the axial resolution of the module. Fig. 10 illustrates the RMS value of z reconstruction versus the source z position. The average value for $\sigma_{zrec}$ is around 3mm.

*4.5 Energy resolution and reconstruction.*

The energy resolution can be deduced from the total amount of photo-electrons collected for each $z_{source}$ coordinate. The spectrum of the sum $N_l(z)+N_r(z)$ obtained for $z_{source}= 0$ is shown in Fig. 11. A Gaussian fit of the photopeak leads to the $\sigma/\mu$ resolution where $\sigma$ and $\mu$ are respectively the RMS and the mean value of the fitted distribution. The best achievable value for the energy resolution 6 % is obtained at the module centre. Fig. 12 shows the simulated energy resolution for the whole module scanned by 5mm steps. It should be stressed that the impact of the PSPMT on the signal resolution is not included (except for the 20 % quantum efficiency).

Since the sum $N_l+N_r$ is not a constant along the z-axis, a correcting factor $f_{cor}(z_{rec})$ has to be applied to get the reconstructed value of the energy $E_{rec}$.

$$E_{rec} = \frac{N_r+N_l}{f_{cor}(z_{rec})} \times 511 \text{ keV} \qquad (8)$$

Using the exponential model for $N_r$ and $N_l$ as described by Eq. (2), it appears that :

$$E_{rec} \propto \sqrt{N_r \cdot N_l} \qquad (9)$$

with

$$\sqrt{N_l(z)\cdot N_r(z)} = A\cdot\frac{N_0}{2}\cdot\exp\left(\frac{-l}{\lambda_{ref}}\right) \quad (10)$$

should be a constant.

However in our Monte Carlo simulation, the energy calculated using Eq. (9) still exhibits a dependency on the z coordinate (see Fig. 13). The variation of the energy value is about 10%. This can be explained as follows: assuming $N_{sc}$ is the total number of UV scintillation photons emitted by an incident 511 keV $\gamma$ photon. Each scintillation photon is tracked in the MC code from the emitting point through the aluminum guide to the photodetectors, taking into account reflection as well as absorption. The total number of reflections by the end of the guide is different from a scintillation photon to another. In (11) $N_i$ is the number of scintillation photons that undergo i reflections.

$$N_{sc} = \sum_{0}^{\infty} N_i \quad (11)$$

If the guide exhibits a reflection coefficient CR, then the total number N of scintillation photons collected at one end of the guide is given by Eq.(12).

$$N = \sum_{0}^{\infty} N_i \times CR^i = \sum_{0}^{\infty} N_i \times \exp[i \times \ln(CR)] \quad (12)$$

Assuming that N can be approximated by an exponential model means that Eq. (12) can be approximated by Eq. (13):

$$N = N_{sc} \times \exp[n_{mean} \times \ln(CR)] = N_{sc} \times CR^{n_{mean}} \quad (13)$$

Where $n_{mean}$ is the weighted average number of reflections the scintillation photons undergo for a given interaction point. In our case this approximation leads to a 10% error on the calculated energy which is very satisfactory. To get a more precise value of the energy deposited by the $\gamma$ photon it is necessary to apply a correction. As a consequence, in a calibration phase, a Gaussian fit of the $N_r(z)+N_l(z)$ distribution is done at each z position. The $[N_r(z)+N_l(z)]_{mean}$ values are stored. This set of points is then fitted using a polynomial expression. Finally, the simulated data are then corrected according to the $z_{rec}$ value previously given by Eq. (7). Fig. 14 shows the result of such a correction on the reconstructed energy, the error is at the level of 1%.

*4.6 Discussion.*

Simulation shows that the exponential model of the light attenuation is realistic. The z reconstructed coordinate can be found using the exponential scheme and the effect of Xenon impurities can be easily introduced in the equations that describe the model. Besides, the main advantage of this model is to qualify the light propagation medium by a single $\lambda_{ref}$ parameter. However, it must be stressed that the exponential approximation induces an error in the energy calculation (10% in our case). Given the very satisfactory results obtained when a polynomial correction is applied, this method has been chosen in the following analysis for the energy reconstruction. On the contrary the exponential model has been kept for the light guide characterization and the z-coordinate reconstruction.

## 5 Experimental results.

As mentioned in section 2 the experimental trigger was conditioned by the coincidence of the LXe cell and the LYSO detector signals. Experimental data analysis was carried out in the same conditions as in the simulation. However a 1mm step scan of the edges of the LXe cell revealed a 4 mm gamma beam width limiting the module performance near its edges. This arose from solid angle calculation once taking into account the source collimation (see section 2). In the present results the scan range of the module was set to +/- 20 mm around the central position. The edges calibration was also used to precisely determine the module centre from LYSO detector point of view. In the following analysis this central position is defined as having z=0. Moreover PSPMT test showed that their response exhibited large fluctuations over the entrance window and that a response mapping was necessary.

### *5.1 Transaxial localisation and PSPMT mapping.*

The PSPMT calibration was carried out using Xenon scintillation light. The first step to perform this mapping is to identify the guide in which the γ interaction took place. The transaxial localisation is derived from a weighted average calculation on the PSPMT anodes signals. The 40 light guide separation is shown in Fig. 15 (the $\sigma_x$ and $\sigma_y$ resolutions are respectively at the level of 1.4 and 0.6 mm). All of the four guides distributed along the x axis can be isolated whereas the collimation performed by the LYSO detector is such that our data are mostly accumulated in the 4 guides located close to the centre along the y axis. The simulation shows that the events reconstructed between two guides are induced by γ rays undergoing Compton scattering in several guides. Such events cannot be processed by PSPMTs. Therefore, in the analysis code, cuts are applied to select events that take place in a single guide. The PSPMTs mapping consists in the determination of a coefficient table that normalize the right and left responses for each guide when the $^{22}Na$ source is located at the module centre.

*5.2 Determination of the attenuation length $\lambda_{ref}$.*

The dynode signals of the two PSPMTs ($N_r$ and $N_l$) were used to determinate the $\lambda_{ref}$ value. The expression $\ln(N_r/N_l)$ is calculated for each position of a 5 mm step scan and then the mean values are derived from gaussian fits of the distributions. The linear behaviour of the result (see Fig. 8) indicates that experimental data can also be processed using an exponential model. This calculation was operated for each of the 16 light guides. The data analysis showed that the 16 $\lambda_{ref}$ values exhibit a 23.9 mm mean value and a 1.0 mm standard deviation. The simulated attenuation length was found to be equal to 26.5 mm for a light guide reflectivity equal to 80 %. Assuming a 78 % reflectivity leads to a simulated $\lambda_{ref}$ equal to 24 mm.

*5.3 The axial resolution.*

The determination of the axial resolution was also based on the dynode signals asymmetry. Data were processed in the same way as in the simulation. The axial resolution is shown in Fig. 10 while the Fig. 16 presents the reconstructed position $z_{rec}$ as a function of the source position $z_{source}$. An axial resolution of 3.0 ± 0.3 mm was obtained since no cuts on the energy were applied. The effects of such cuts will be presented in section 5.6.

*5.4 The energy resolution and reconstruction.*

The energy resolution is derived from the fit of the photopeak in the $N_r$+$N_l$ spectrum. The spectrum corresponding to $z_{source}$=0 is shown in Fig. 17. The energy resolution for the whole module is shown in Fig. 12. The best energy resolution achievable is about 10 % near the centre position of the module. The energy calculated using Eq. (9) also exhibits a 10 % error. The energy correction using a polynomial fit reduces this error to 3 %.

*5.5 Time resolution.*

The time resolution of the module prototype has been measured using the TDC signal. Fig. 18 summarizes the time resolution measurement along the module by step of 5 mm. A mean value of about 700 ps has been found when no energy cut is applied. It is improved up to 550 ps with an energy cut of 400 keV.

*5.6 Summary of the module performances and discussion.*

The optimization of the studied Liquid Xenon PET prototype depends on three relevant parameters:

- the energy resolution at the 511 keV photo-peak that permits to cut down photons having undergone Compton scattering,
- the time response that helps to level down random coincidence rate,
- the DOI measurement that makes the system insensitive to any parallax effect.

Concerning the first two relevant parameters (see Tab. 2), the tested prototype exhibits a 10% (RMS) energy resolution and a 550 ps time resolution. Better energy and time resolutions can be achieved using LXe [35,36] but to optimize these resolutions it is necessary to minimize the amount of UV photons absorbed before they reach a photo detector. In Ref. [36], detectors use ultra pure xenon (on line purification is operated) and high reflectivity material such as Teflon, which would not be appropriate in the present application where specular reflection is requested. Moreover in Ref. [35,36] the photodetectors are immersed in the LXe to improve the light collection efficiency. On the contrary, the DOI measurement described in the present paper is straightforward, since determined by the number of pixels in the module, whereas the z-coordinate relies on UV photon absorption by the detector itself. For instance, a 90% reflectivity is less favorable than 80%. Consequently the optimal values for z, time and energy resolutions are a matter of compromise. Furthermore, no purification of the Liquid Xenon is operated in the perspective of simple experiment handling to satisfy future biology application. Simulation shows that increasing $\lambda_{ref}$ levels up the number of photoelectrons collected. By doing so, better $\sigma_E/E$ and $\sigma_t$ resolutions can be achieved but to the detriment of $\sigma_z$. Nevertheless, if the number of photo electrons collected is increased by using PSPMTs with higher quantum efficiency or if the present PSPMTs are immersed in the LXe, then better $\sigma_E/E$, $\sigma_t$ and $\sigma_z$. resolutions are reachable. The present 550 ps (RMS) time resolution would not permit to make the proposed LXe PET prototype a Time-of-Flight PET (TOF-PET) but would make possible to access on-line to the bio distribution of a medical tracer in laboratory animals (drug developments).

Concerning the DOI measurement, it has been demonstrated that a 400 keV energy cut enables to reach a resolution on z of 2.5 ± 0.2 mm (see Tab. 2) which is a promising result when compared to [34]. A simulation of the present test bench with PSPMTs that time immersed in the Liquid Xenon shows that a $\sigma_z$ at the level of 1.5 mm is reachable.

On the contrary the $\sigma_x$ and $\sigma_y$ measurements are already very competitive. In our detector scheme, the transverse coordinates x and y of the interaction point are derived from the address of the hit guide. With a matrix of 40 light guides each featuring a 5 x 2 mm$^2$ cross section, the $\sigma_x$ and $\sigma_y$ resolutions are respectively at the level of 1.4 and 0.6 mm. A second matrix of 100 light guides each featuring a 2x2 mm$^2$ is currently in test thus exhibiting a $\sigma_x$ and $\sigma_y$ resolution of 0.6 mm. Such a resolution would make possible, with the proposed LXe μPET, to image the distribution of molecular probes at sub-millimeter resolution in laboratory animals.

## 6 Conclusion.

A first prototype module of a parallax free μPET using LXe in an axial geometry has been built. A module is optically subdivided into 40, 5 x 2 $mm^2$, $MgF_2$-coated aluminum UV light guides read on each end by PSPMTs. The light guide modeling has been extensively discussed. The exponential model chosen has been found to be very attractive since light propagation can be described with simple analytic expression. Furthermore it allows the LXe pollution to be easily accounted for. An experimental test bench has been built to measure the transverse localization of an interaction in the module, i.e. x and y measurements, but also to evaluate the axial localization (z coordinate), the energy and time resolutions. A Monte-Carlo simulation of this test bench has been carried out using the GATE-GEANT4 package. Both experimental and simulated data have been analyzed using the exponential scheme. The agreement is very satisfactory. The resolution on the z coordinate has been found to be 2.5 mm once the photopeak is isolated with a cut on the reconstructed energy at 400 keV, the best achievable energy resolution is 10 % on the data and time resolution is about 550 ps. These results are promising. Better resolutions may be achieved by immersing the PSPMTs in the LXe, this solution is envisaged in a near future. The simulation of the whole scanner is underway to quantify its performance on reconstructed images.

**Acknowledgments.**

Financial support from Institut National de Physique Nucléaire et de Physique des Particules (IN2P3), Région Rhône-Alpes and Université Joseph Fourier is acknowledged. Yves Carcagno from the LPSC laboratory is thanked for his help in handling the cryogenic part of the experimental test bench. We are grateful to Eric Perbet from the LPSC laboratory for his implication in the technical drawings of the mechanical part of the experimental test bench.

**References.**

[1] S. Weber and A. Bauer European. Journal of Nuclear Medicine and Molecular Imaging **31** (2004) 1545.

[2] Y. Yang *et al.*, Phys. Med. Biol. **49** (2004) 2527.

[3] K. Wienhard *et al.*, IEEE Trans. Nucl. Sci. **49** (2001) 104.

[4] Y.C. Tai and D.F. Newport, Journ. of Nucl. Medicine **46** (2005) 3.

[5] S. Jan, *private communication*, measurement done at CEA/SHFJ.

[6] S. Kubota *et al.*, Phys. Rev. **B 20** (1979) 3486.

[7] S. Kubota *et al.*, Nucl. Instr. and Meth. **A 196** (1982) 101.

[8] T. Doke *et al.,* Japan. J. Appl. Phys. **41** (2002) 1538.

[9] The Crystal Clear Collaboration : http://crystalclear.web.cern.ch/

[10] S. Eidelman *et al.,* (Particle Data Group), Phys. Lett. **B 592**(2004) 1.

[11] V. Chepel *et al.*, Nucl. Instr. and Meth **A 392** (1997) 427, IEEE Trans. Nucl. Sci. 46 (1999) 1038.

[12] J. Collot *et al.*, Proc. of the IXth Intern. Conference on Calorimetry in High Energy Physics (CALOR 2000).Oct. 2000, Annecy (France), Eds. B. Aubert *et al*. (Frascati Physics Series Vol 21), pp. 305

[13] S. Jan *et al.*, Proc. of International Conference Imaging Technologies in Biomedical Sciences (ITBS 2001), May 2001, Milos Island (Greece).

[14] S. Jan, PhD Thesis, Université Joseph Fourier (Grenoble, France), Sept. 2002.

[15] F. Nishikido *et al.,* Japanese Journal of Applied Physics **43** (2004) 779.

[16] G.J. Davies *et al.*, Phys. Lett. **B320** (1994) 395.

[17] E. Aprile *et al.*, Nucl. Instr. and Meth. **A 461** (2001) 256.

[18] R. Bernabei *et al.*, Nucl. Instr. and Meth. **A 482** (2002) 728.

[19] A. Incicchitti *et al.*, Nucl. Instr. and Meth. A **289** (1990) 236.

[20] S. Agostinelli *et al.*, Nucl. Instr. and Meth. A **506** (2003) 250.

[21] S. Jan *et al.*, Phys. Med. Biol. **49** (2004) 4543-4561.

[22] J. Seguinot *et al*, Nucl. Instr. and Meth. **A 354** (1995) 280.

[23] V.N. Solovov *et al.*, Nucl. Instr. and Meth. **A 516** (2004) 462.

[24] Hamamatsu Photonics, 8 Rue du Saule Trapu, Parc du Moulin de Massy, 91300 Massy, France.

[25] S. Jan *et al.*, GATE User's Guide.

[26] S. Jan et al., IEEE Trans. Nucl. Sci. **52** (2005) 102.

[27] K.S Shah *et al.*, IEEE Trans. Nucl. Sci. **51** (2004) 91.

[28] E. Gramsch *et al.,* IEEE Trans. Nucl. Sci. **50** (2003) 307.

[29] K.C. Burr *et al.*, Nuclear Science Symposium Conference Record, 2003 IEEE Volume 2, Issue , 19-25 Oct. 2003 Page(s): 877 - 881 Vol.2.

[30] A. Braem *et al.*, Nucl. Instr. and Meth. **A 571** (2007) 131.

[31] A. Braem *et al.*, Nucl. Instr. and Meth. **A 320** (1992) 228.

[32] N. Ishida *et al.*, Nucl. Instr. and Meth. **A 327** (1993) 152.

[33] Baldini *et al.*, Nucl. Instr. And Meth. **A 545** (2005) 753.

[34] I. Vivaldi *et al.*, Nucl. Instr. and Meth. **A 564** (2006) 506.

[35] T. Doke *et al.*, Nucl. Instr. and Meth. **A 569** (2006) 863.

[36] K. Ni *et al*, JINST **1** (2006) 9004.

**List of Figures.**

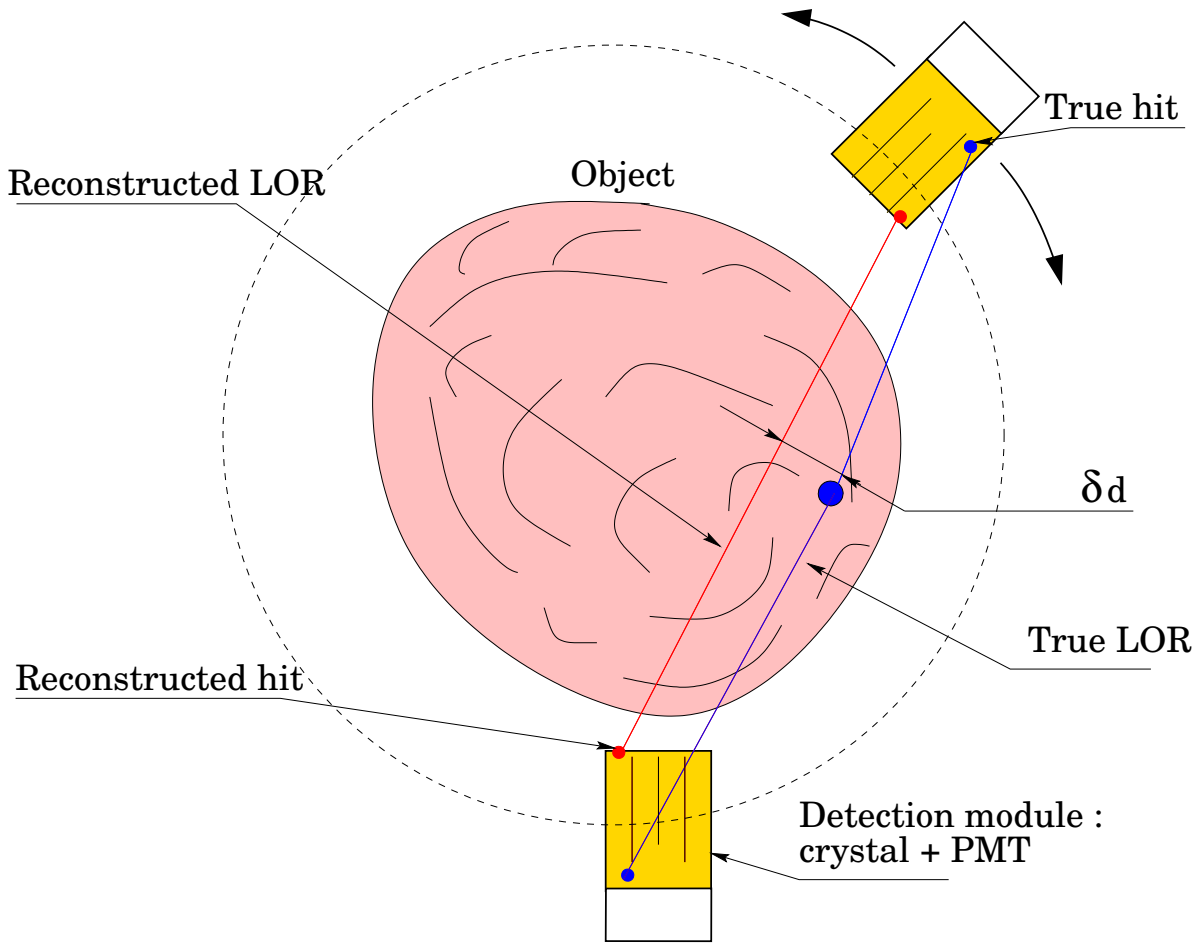

Fig.1. Parallax effect due to the non-measurement of the depth of interaction.

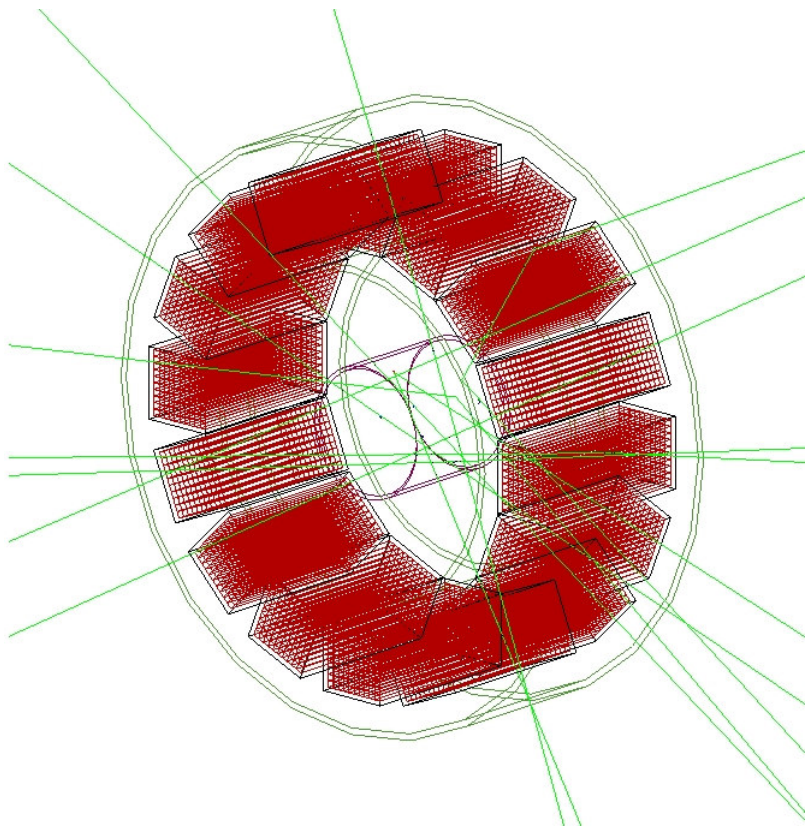

Fig. 2. The LXe μPET geometry.

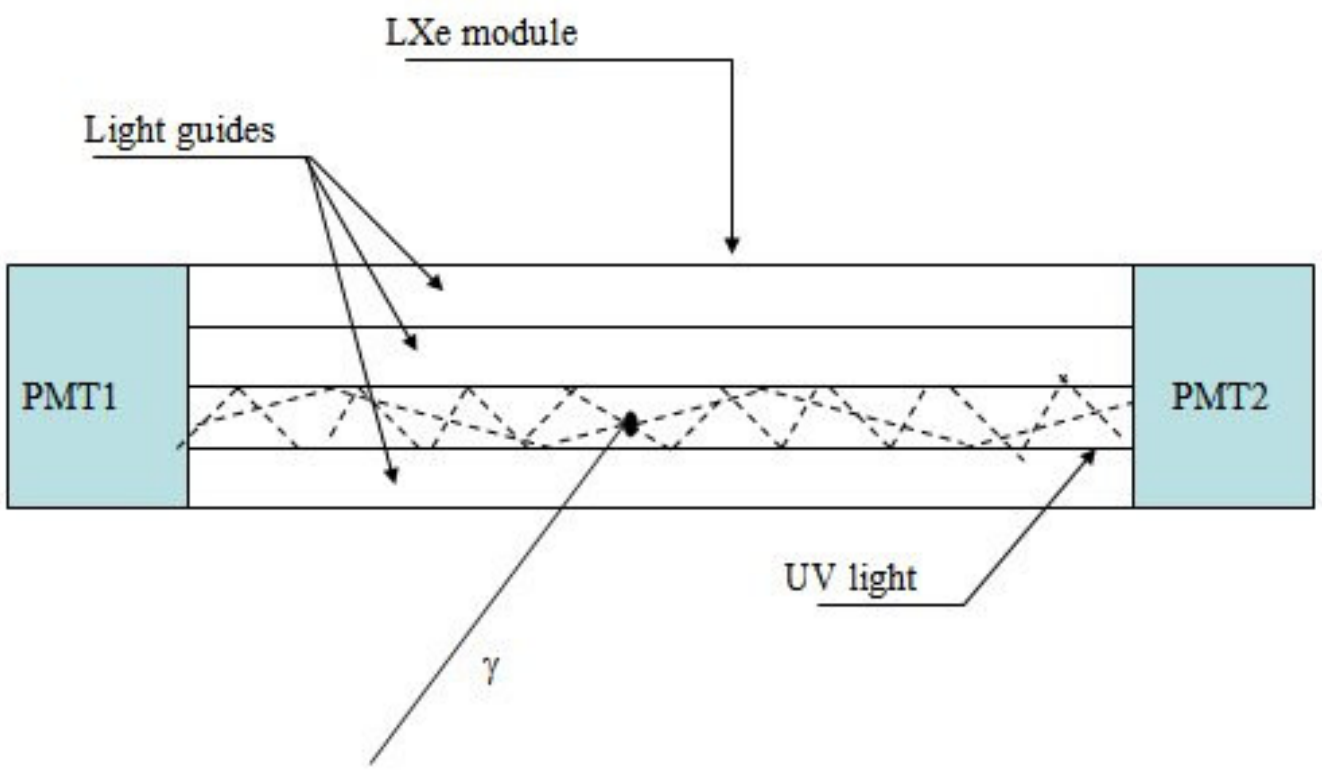

Fig. 3. Sketch of an elementary module of the LXE μPET camera : the z-axis is along the axial direction of the μPET.

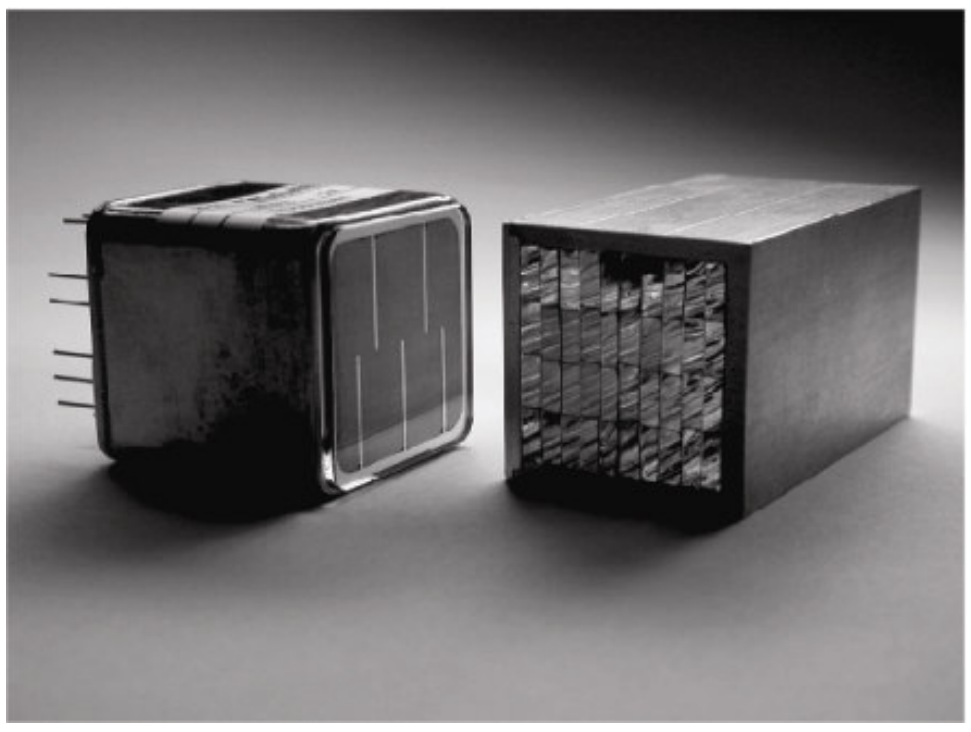

Fig. 4. Hamamatsu [24] R8520-06-C12 Position Sensitive Photomultiplier tubes and the 40, 2x5 $mm^2$ Mg-F2 coated aluminum VUV light guides.

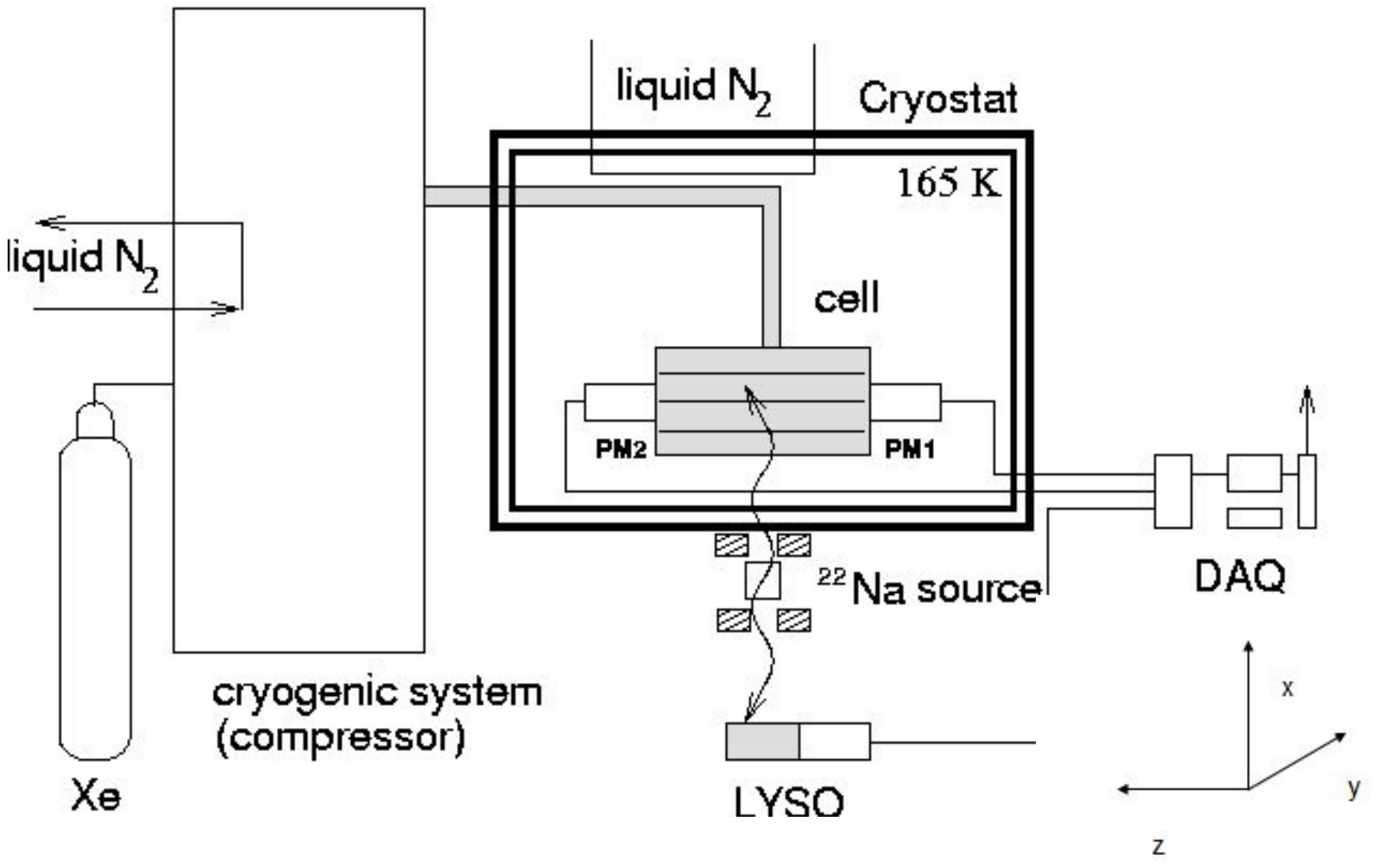

Fig. 5. The experimental set-up used for the test of the prototype module.

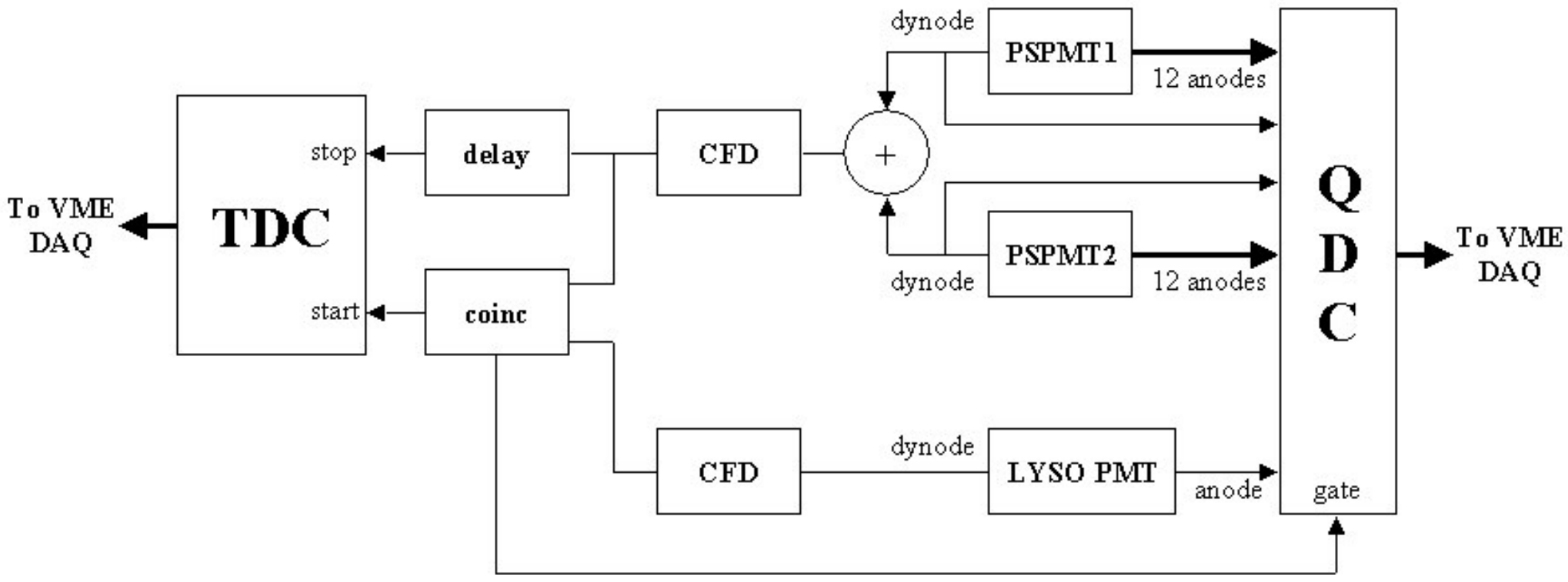

Fig. 6. Data Acquisition System.

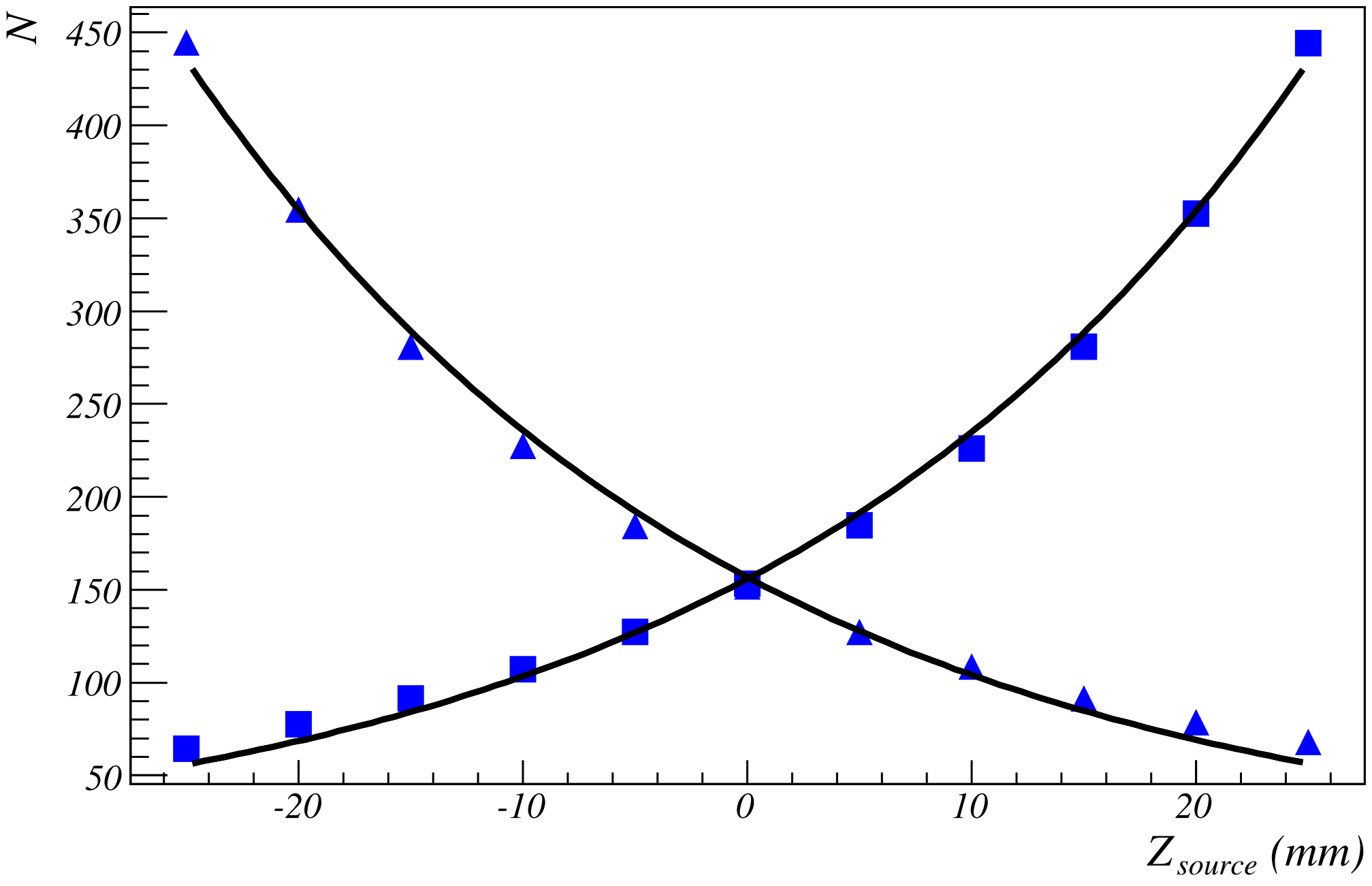

Fig. 7. Number N of photoelectrons collected at each module end ($N_r$ = square , $N_l$ = triangle) versus the $z_{source}$ position (the error bars are smaller than the marker size) - simulation result.

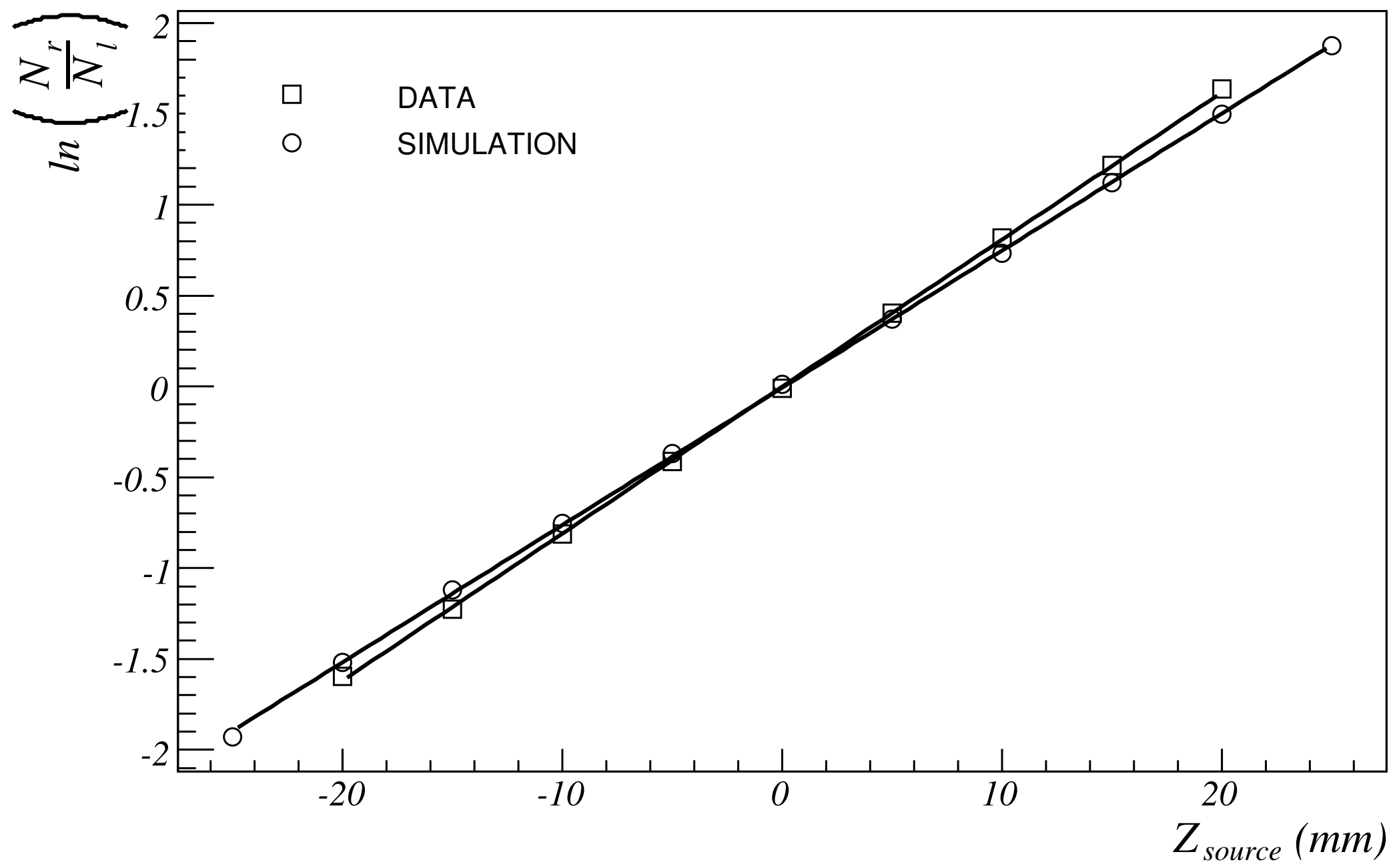

Fig. 8. $\ln(N_r/N_l)$ versus $z_{source}$ position (the error bars are smaller than the marker size).

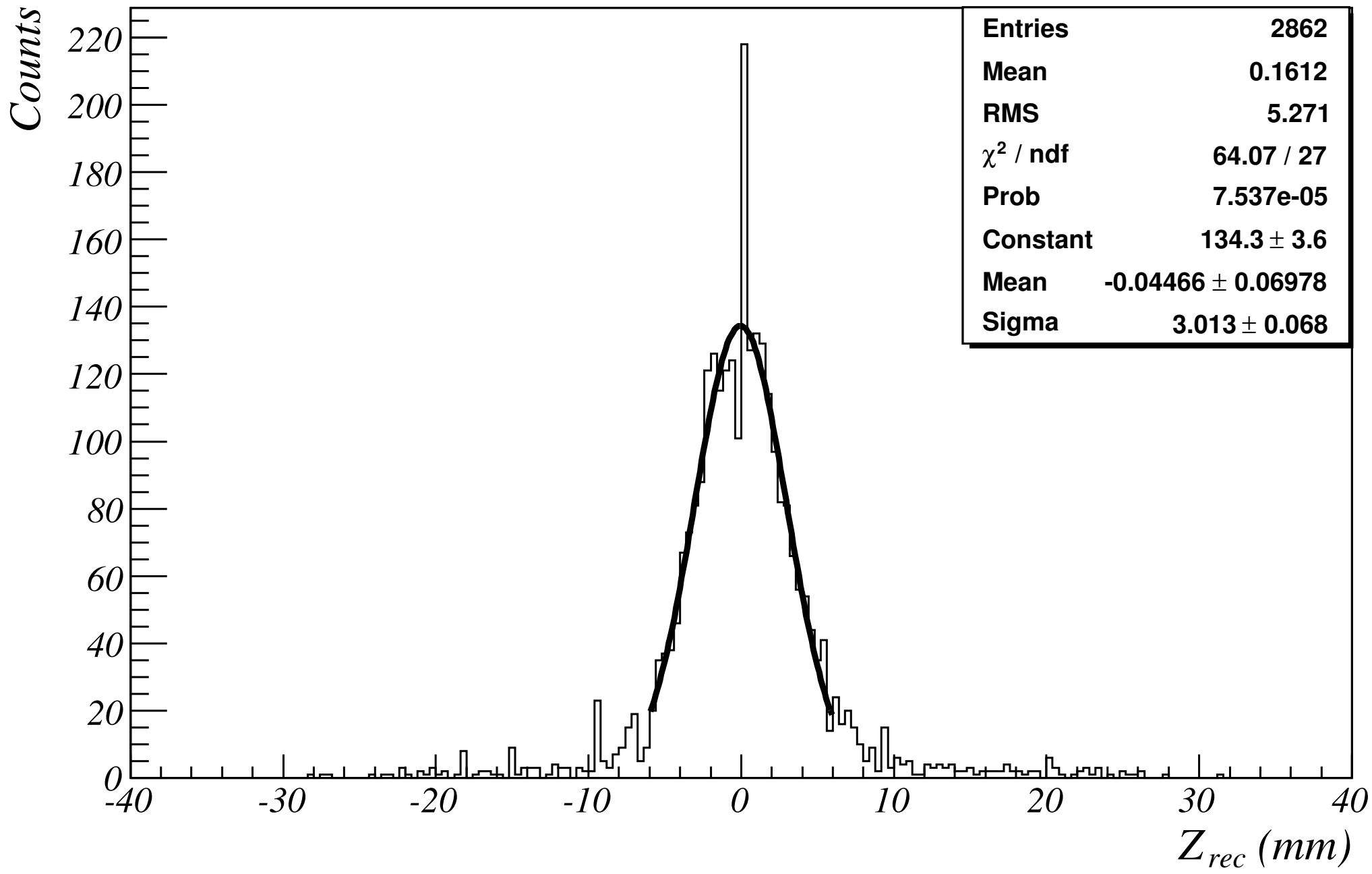

Fig. 9. Distribution of the reconstructed $z_{rec}$ position for a source located at $z_{source}$=0 - simulation result.

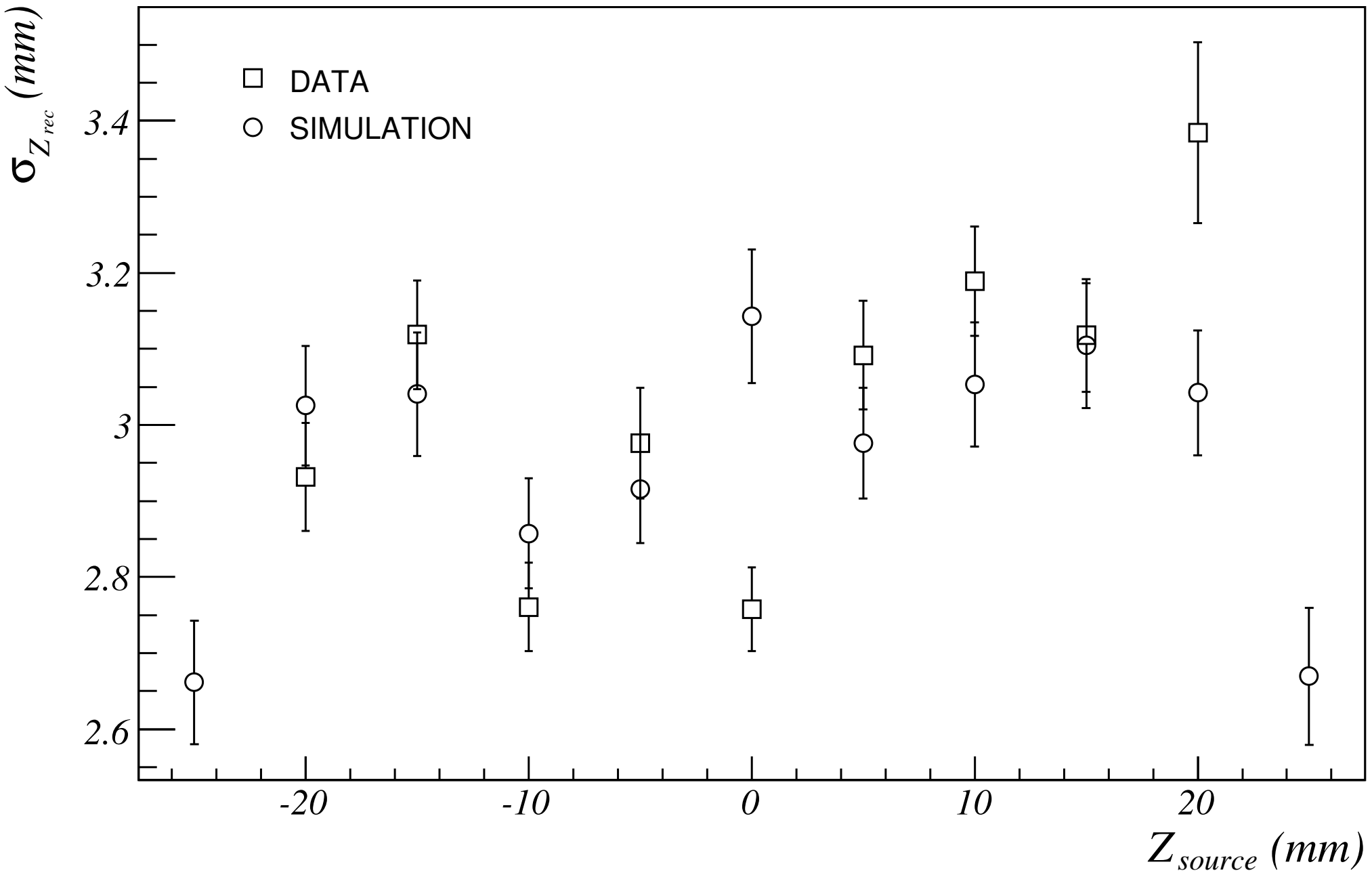

Fig. 10. RMS value of $z_{rec}$ versus the $z_{source}$ position.

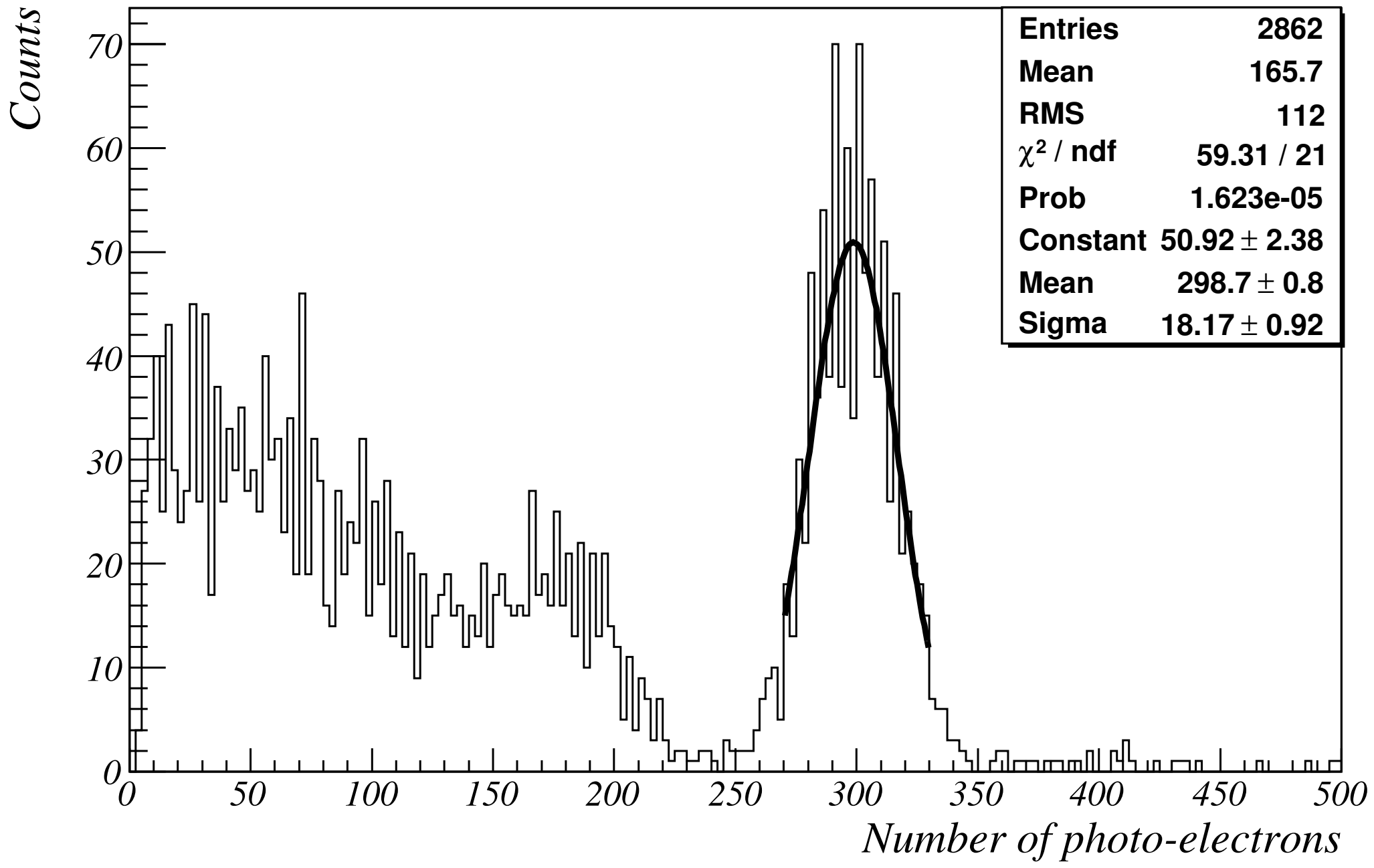

Fig. 11. Spectrum of the sum $N_l(z)+N_r(z)$ obtained for $z_{source}=0$ in a single 2x5 $mm^2$ guide – simulation result.

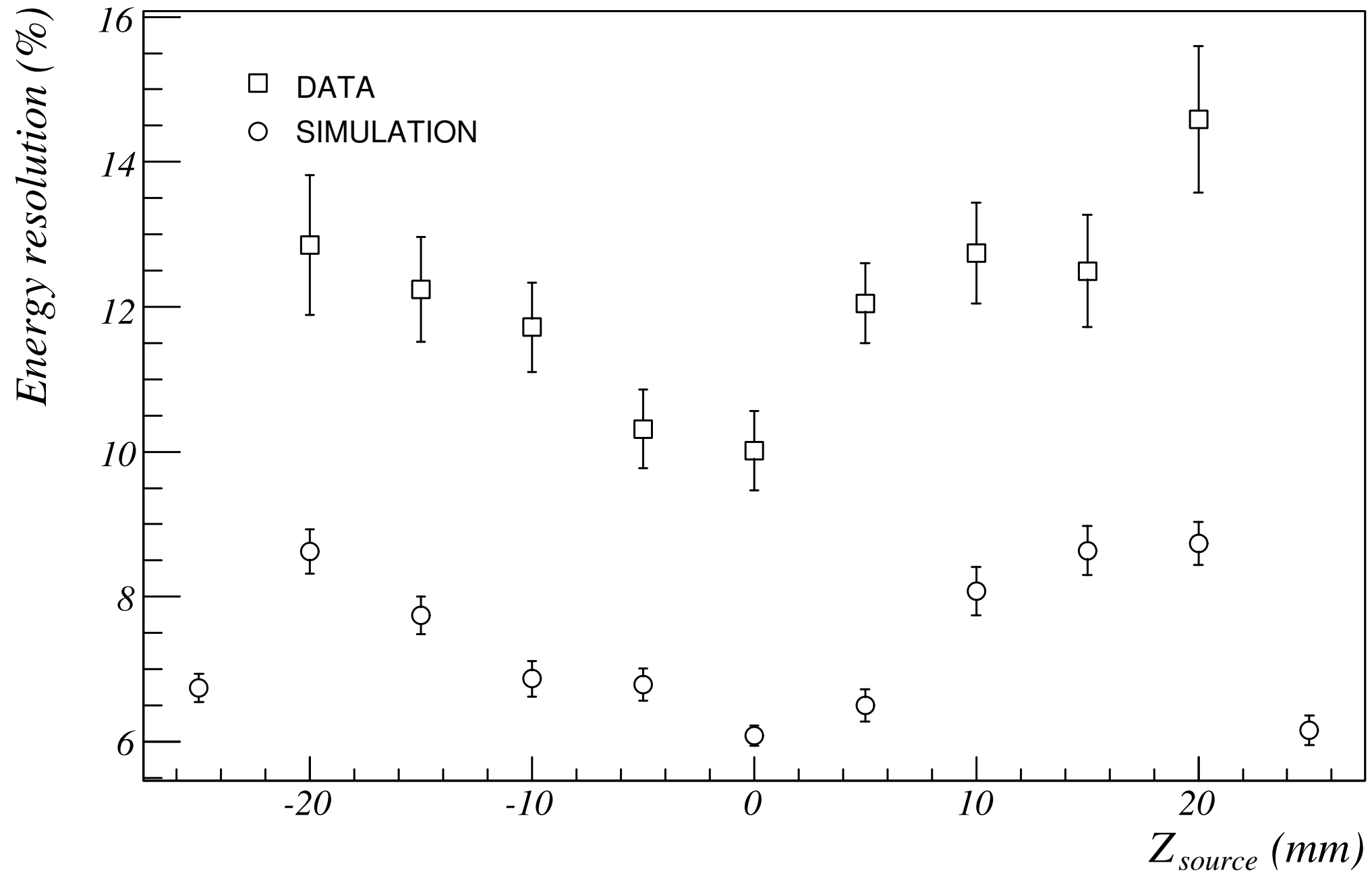

Fig. 12. Energy resolution for the whole module scanned by 5mm steps.

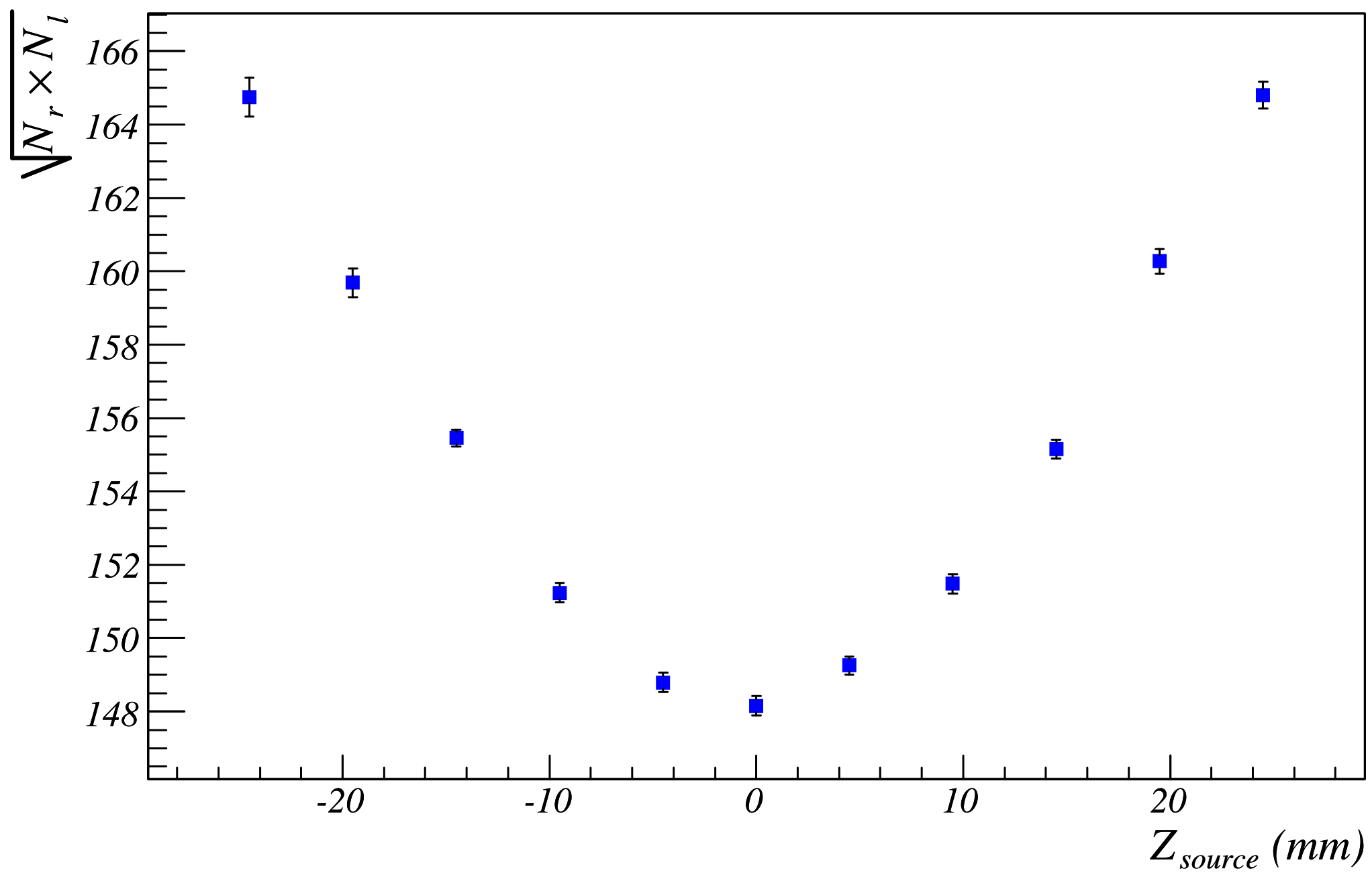

Fig. 13. The $\sqrt{N_r \times N_l}$ distribution versus the $z_{source}$ position – simulation result.

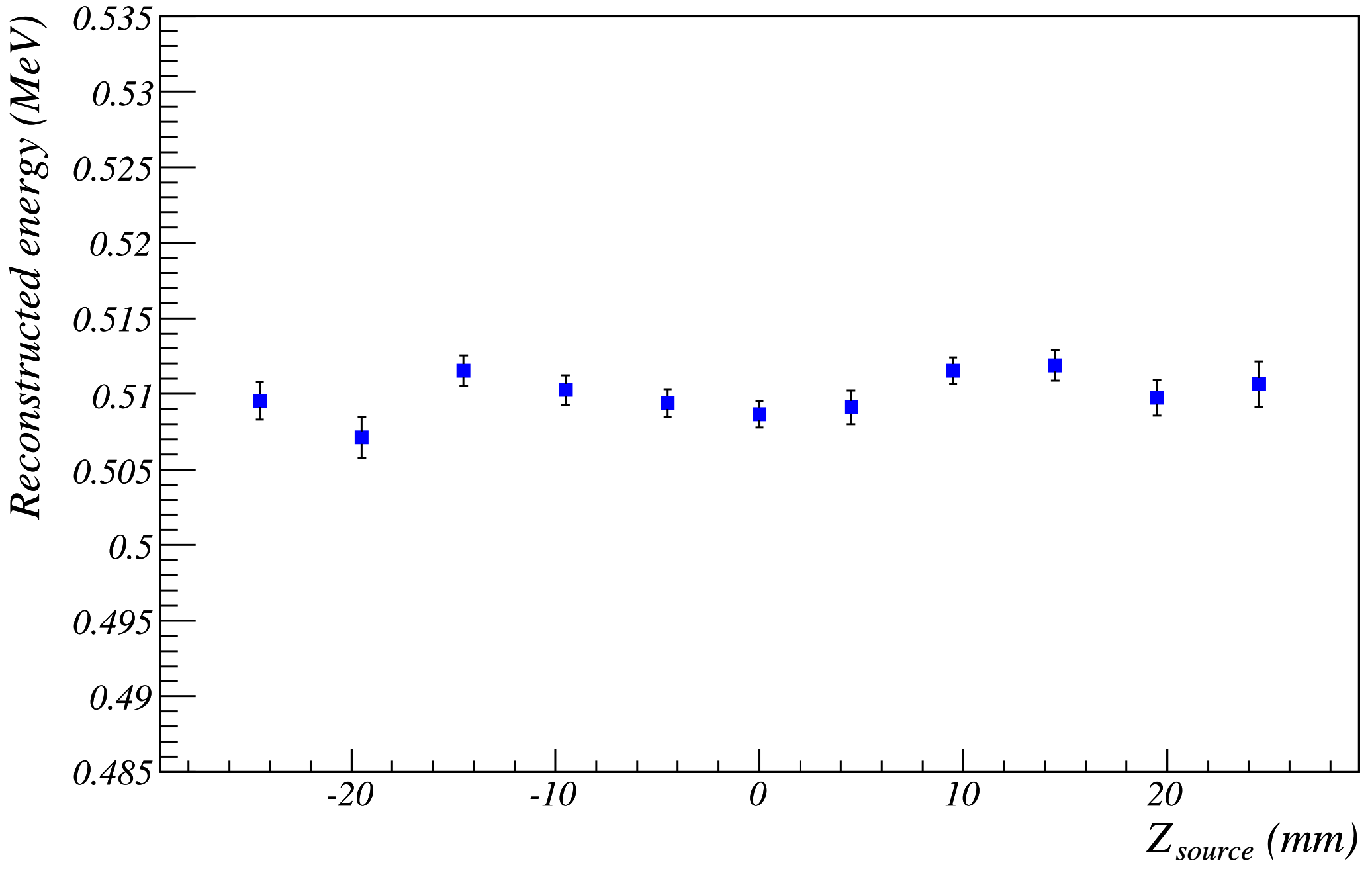

Fig. 14. Reconstructed energy versus $z_{source}$ position – simulation result.

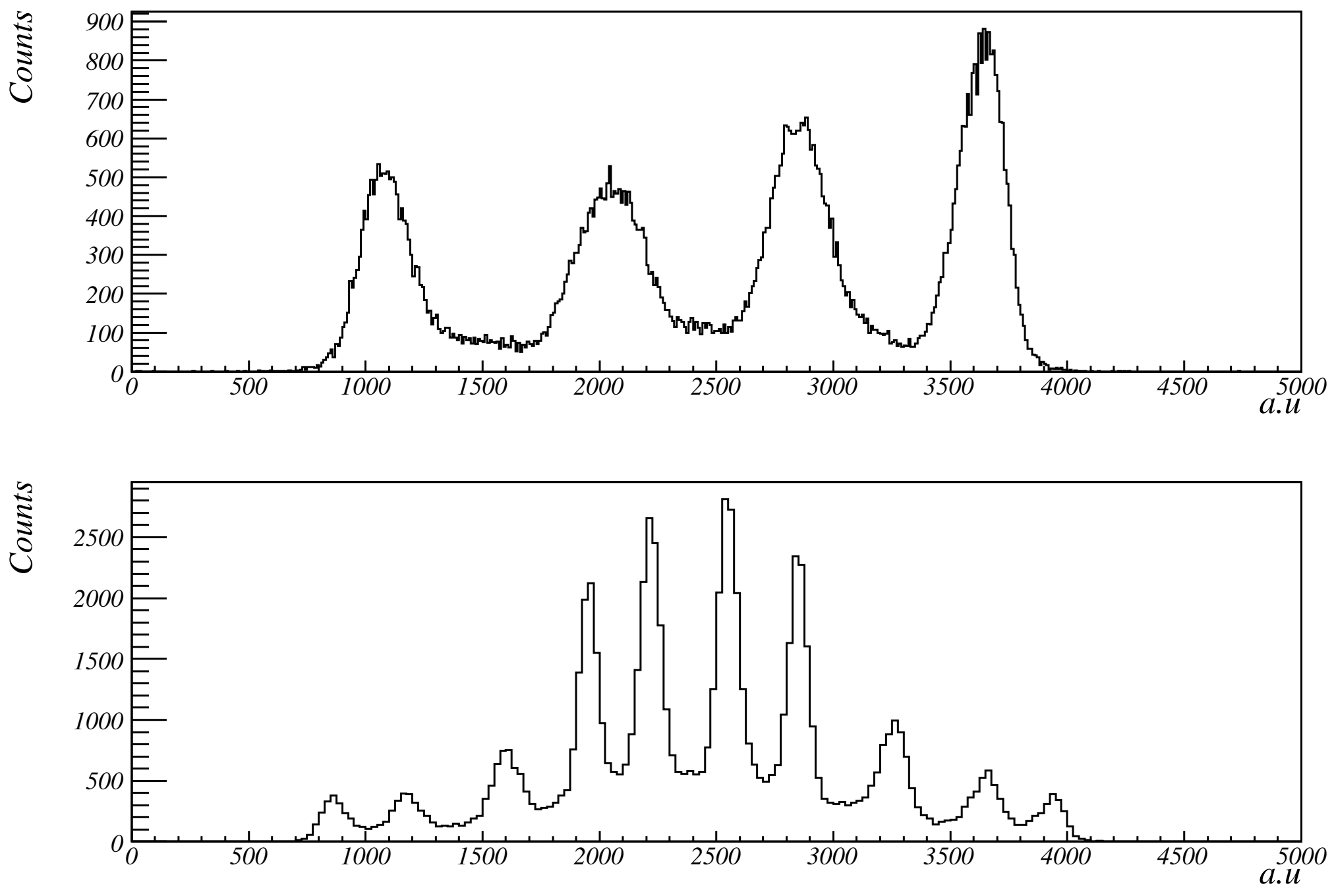

Fig. 15. The x (top) and y (bottom) barycentre distributions computed by the data acquisition system and derived from the PSPMT anode signals.

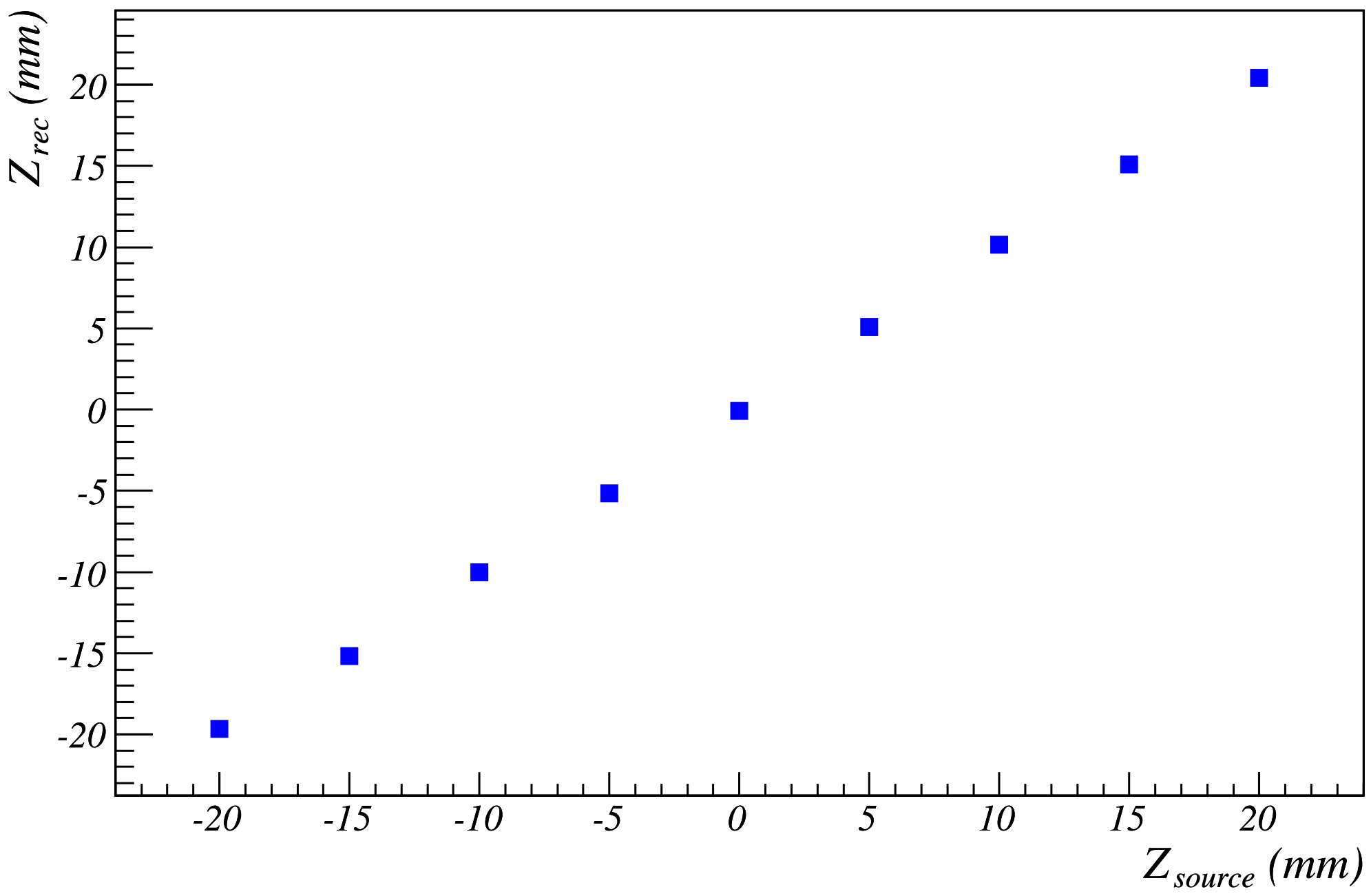

Fig. 16. Distribution of $z_{rec}$ versus $z_{source}$ for the experimental data (the error bars are smaller than the marker size).

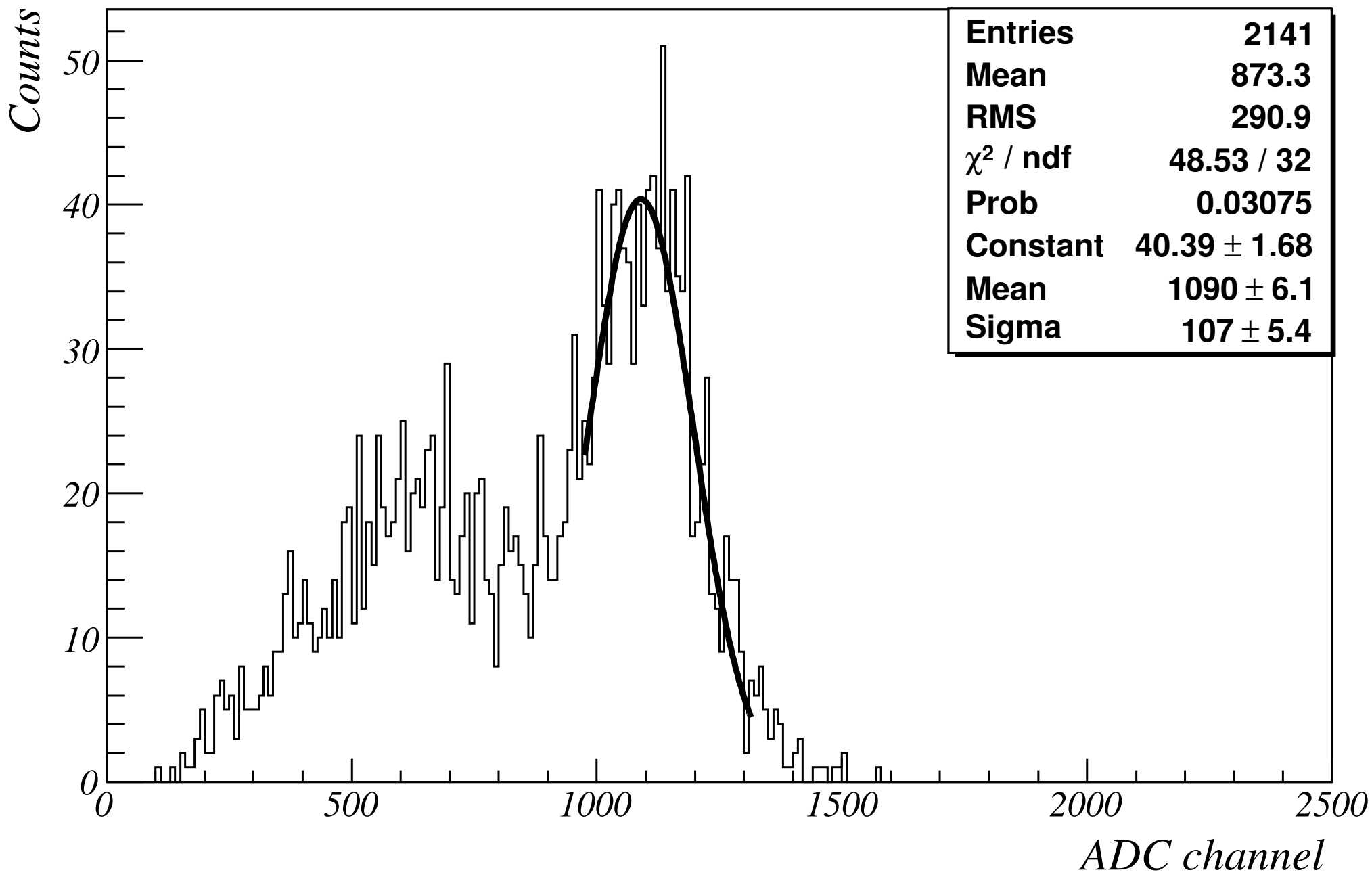

Fig. 17. The spectrum of the sum $N_l(z)+N_r(z)$ obtained for $z_{source}=0$ in a single guide for the experimental data.

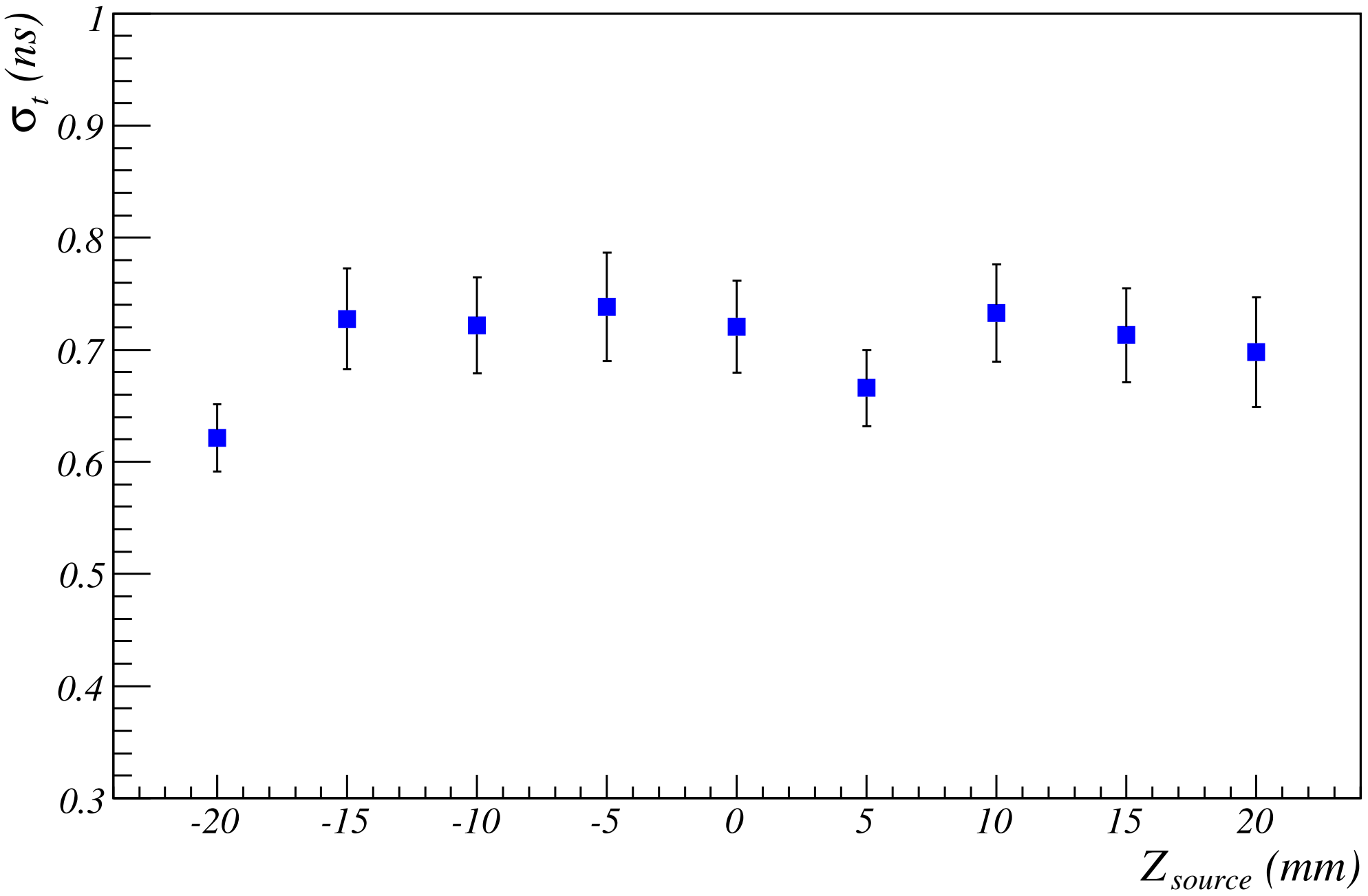

Fig. 18. The time resolution scanned over the whole module by 5 mm steps - experimental data analysis.

**List of tables.**

| Scintillator | Density (g $cm^{-3}$) | Decay time (ns) | Light Yield (UV/MeV) |
|---|---|---|---|
| LXe | 3.06 | 3-30 [6,7] | 46000 [8] |
| NaI | 3.67 | 230 | 38000 |
| BGO | 7.13 | 300 | 9000 |
| LSO | 7.4 | 40 | 27000 |

Table 1 : Comparison of physical properties of LXe and commonly used crystals. Values from [9,10], unless otherwise stated.

| | Experimental data (selection criterion) | Simulation (no PMTs) (selection criterion) |
|---|---|---|
| λ | 24mm | 24mm<br>CR=0.78 et LA=38° |
| Energy resolution | 10% | 6% |
| Z resolution (RMS) | 2.5mm (400keV)<br>3mm | 2.5mm (400keV)<br>3mm |
| Time resolution (RMS) | 550ps (400keV)<br>700ps | |

Table 2 : Comparison of experimental results and simulation - summary table.